\documentstyle[preprint,eqsecnum,aps,epsf]{revtex}
\newif\iftightenlines\tightenlinesfalse
\tightenlines\tightenlinestrue
\def\eslt{E\llap/_T}
\def\to{\rightarrow}

\def\tb{\tilde b}

\def\tst{\tilde t}
\def\ttau{\tilde \tau}

\def\tg{\tilde g}
\def\tnu{\tilde\nu}
\def\tell{\tilde\ell}
\def\tq{\tilde q}
\def\tw{\widetilde W}
\def\tz{\widetilde Z}
\begin{document}
\draft
\preprint{\vbox{\baselineskip=14pt%
   \rightline{FSU-HEP-980730}\break 
   \rightline{IFT-P.059/98}\break 
   \rightline{BNL-HET-98/25}\break 
   \rightline{UH-511-914-98}
}}
\title{PROBING MINIMAL SUPERGRAVITY AT THE \\
CERN LHC FOR LARGE $\tan\beta$}
\author{Howard Baer$^1$, Chih-Hao Chen$^1$, Manuel Drees$^2$, 
Frank Paige$^3$\\
 and Xerxes Tata$^{4}$}
\address{
$^1$Department of Physics,
Florida State University,
Tallahassee, FL 32306 USA
}
\address{
$^2$IFT, Univ. Estadual Paulista, 01405--900 S\~ao Paulo, Brazil
}
\address{
$^3$Brookhaven National Laboratory, 
Upton, NY 11973 USA
}
\address{
$^4$Department of Physics and Astronomy,
University of Hawaii,
Honolulu, HI 96822 USA
}
\date{\today}
\maketitle
\begin{abstract}

For large values of the minimal supergravity model parameter 
$\tan\beta$, the tau lepton and the bottom quark Yukawa couplings become
large, leading to reduced masses of $\tau$-sleptons and $b$-squarks
relative to their first and second generation counterparts, and to
enhanced decays of charginos and neutralinos to $\tau$-leptons and
$b$-quarks. We evaluate the reach of the CERN LHC $pp$ collider for
supersymmetry in the mSUGRA model parameter space. We find that values
of $m_{\tg}\sim 1500-2000$ GeV can be probed with just 10 fb$^{-1}$ 
of integrated luminosity for $\tan\beta$ values as high as 45, so that mSUGRA
cannot escape the scrutiny of LHC experiments by virtue of having
a large value of $\tan\beta$. We also perform a case study of an mSUGRA
model at $\tan\beta =45$ where $\tz_2\to \tau\ttau_1$ and 
$\tw_1\to \ttau_1\nu_\tau$ with $\sim 100\%$ branching fraction. In this case,
at least within our simplistic study, we show that a di-tau mass edge, 
which determines the value of $m_{\tz_2}-m_{\tz_1}$, can still
be reconstructed. This 
information can be used as a starting point for reconstructing SUSY
cascade decays on an event-by-event basis, and can provide a strong constraint
in determining the underlying model parameters.
Finally, we show that for large $\tan\beta$ there can be an observable
excess of $\tau$ leptons, and argue that $\tau$ signals might serve to provide
new information about the underlying model framework.
\end{abstract}

\medskip

\pacs{PACS numbers: 14.80.Ly, 13.85.Qk, 11.30.Pb}


\section{Introduction}

The minimal supergravity model (mSUGRA)~\cite{sugra} provides a
well-motivated and economical realization the Minimal
Supersymmetric Standard Model\cite{mssm}, or MSSM.  In mSUGRA,
supersymmetry is broken in the ``hidden sector'' which consists of
fields which couple to the fields of the MSSM only
gravitationally. Thus, SUSY breaking is communicated to the visible
sector MSSM fields via interactions of gravitational strength.  The
technical assumption of minimality implies that kinetic terms for
matter fields take the canonical form; this assumption, which is
equivalent to assuming an approximate global $U(n)$ symmetry between
$n$ chiral multiplets, leads to a common mass squared $m_0^2$ (defined
at a high scale $M_X \sim M_{GUT}-M_{Planck}$) for all
scalar fields, and a common trilinear scalar coupling $A_0$ for all $A$
parameters. These parameters, which determine the sparticle-particle
mass splitting in the observable sector, are taken to be of similar magnitude
as the weak scale, $M_{weak}$. In addition, motivated by the apparently
successful gauge coupling unification in the MSSM, one usually adopts
a common value $m_{1/2}$ for all gaugino masses at the scale
$M_{GUT}\simeq 2\times 10^{16}$ GeV. For simplicity, it is commonly
assumed that in fact the scalar masses and trilinear terms unify at
$M_{GUT}$ as well. The resulting effective theory, valid at energy
scales $E<M_{GUT}$, is then just the MSSM with the usual soft SUSY
breaking terms unified at $M_{GUT}$.  The soft
SUSY breaking scalar and gaugino mases, the trilinear $A$ terms and in
addition a bilinear soft term $B$, the gauge and Yukawa couplings and
the supersymmetric $\mu$ term are all then evolved from $M_{GUT}$ to
some scale $M\simeq M_{weak}$ using renormalization group equations
(RGE's). The large top quark Yukawa coupling causes the squared mass
parameter of one of the Higgs fields $H_u$ to be reduced. For
phenomenologically viable choices of
parameters, $m_{H_u}^2$ becomes negative so that electroweak symmetry is
spontaneously broken, and $\mu^2$ can be
determined in terms of $M_Z^2$.
It is customary to trade the parameter $B$ for
$\tan\beta$, the ratio of Higgs field vacuum expectation
values. Finally, it is assumed that superpotential interactions conserve
$R$-parity.  The resulting weak
scale spectrum of superpartners and their couplings can thus be
derived in terms of four continuous plus one discrete parameters
\begin{equation}
m_0,\ m_{1/2},\ A_0,\ \tan\beta,\ {\rm and} \mathop{\rm sgn}(\mu),
\end{equation}
in addition to the usual parameters of the Standard Model (SM).

The implications of the mSUGRA model for supersymmetry searches at the
CERN Large Hadron Collider (LHC) have been examined by several groups
in Refs.~\cite{tdrs,baer,snowmass,hinchliffe,cms}. In Refs.~\cite{baer} the
reach of the LHC for SUSY via a variety of search channels has been
obtained. In channels with jets plus missing transverse momentum
$\eslt$ plus 0-3 isolated leptons, values of $m_{\tg}\sim
1500-2000$ GeV could be probed with just 10 fb$^{-1}$ of integrated
luminosity. This compares well with the reach obtained CMS and ATLAS
studies \cite{tdrs} using somewhat different strategies to isolate the signal.
In addition, for part of this parameter space, a
characteristic edge~\cite{bcpt} in the dilepton mass spectrum gave precision
information on the mass difference $m_{\tz_2}-m_{\tz_1}$. This
measurement could be used as a starting point in various ``case
studies'' to ultimately reconstruct many of the sparticle masses in
cascade decay chains\cite{snowmass,hinchliffe}. These studies, using
the event generator ISAJET\cite{isajet}, were at the time limited to
low values of the parameter $\tan\beta \alt 10$.

For higher values of the parameter $\tan\beta$, the $b$-quark and $\tau$-lepton
Yukawa couplings can become large. This can affect SUSY 
phenomenology in several ways.
\begin{itemize}
\item Large $b$ and $\tau$ Yukawa couplings cause the
$m_{\ttau_{L/R}}^2$ and $m_{\tb_{L/R}}^2$ soft SUSY breaking masses to
run to weak scale values that are lower than the corresponding mass
terms for first and second generation squarks and sleptons. Also, for
large values of $\tan\beta$, there can be large off-diagonal mixing in
sbottom and stau mass matrices. Together, these effects can make the
physical stau and sbottom mass eigenvalues significantly lower than
their first and second generation squark and slepton counterparts
\cite{dn}. 
\item Contributions to the RGE proportional to the squared $b$ and
$\tau$ Yukawa couplings reduce the mass of the $CP$--odd Higgs boson
$m_A$, which in turn reduces the masses of the $CP$--even Higgs boson
$H$ and of the charged Higgs boson \cite{dn}. The upper bound on $\tan
\beta$ is often determined by the experimental lower bound on $m_A$.
\item The relatively lower stau and sbottom masses result in larger 
production cross sections for third generation SUSY particles compared to
first and second generation SUSY particles.
\item The large $\tau$ and $b$ Yukawa couplings, along with relatively
light values of $\tb_i$ and $\ttau_i$ masses can yield enhanced decays
of gluinos to $b$-quarks \cite{dn,bartl} and of charginos and neutralinos
to $\tau$ leptons and $b$ quarks \cite{ltanb,ltanbtev}.

\end{itemize}
As a result, for large values of $\tan\beta$, we expect SUSY events to contain
many more $b$-quarks and $\tau$-leptons than anticipated in earlier
studies that were mostly carried out for relatively
low values of $\tan\beta$. Recently, a number of improvements
\cite{ltanb} have been made to the event generator ISAJET
\cite{isajet} to allow realistic event generation in supersymmetric
models even if $\tan\beta$ is large.

The consequences of the mSUGRA model at large $\tan\beta$ for the
Fermilab Tevatron collider have been examined in
Ref.~\cite{ltanbtev}. In this study, it was found that for large
$\tan\beta$, significantly fewer hard isolated $e$'s and $\mu$'s are
produced in SUSY events, so that the reach for SUSY in various isolated
lepton channels is greatly {\it reduced} compared to the corresponding
reach at low $\tan\beta$. For instance, much of the reach of the
Fermilab Main Injector (and possible luminosity upgrades thereof)
comes from $\tw_1\tz_2\to 3\ell$
production \cite{trilep}, where $\ell =e$ or $\mu$, if $\tan\beta$ is small.
However, for high $\tan\beta$ values,
decays such as $\tw_1\to \tau\nu\tz_1$ and $\tz_2\to\tau\bar{\tau}\tz_1$
can become dominant so that far fewer hard isolated leptons are
produced, and the reach for SUSY is considerably diminished. In fact, in
Ref.~\cite{ltanbtev}, it was found that for $\tan\beta =45$, there would
be {\it no reach} of the Fermilab Tevatron Main Injector ($\int{\cal
L}dt=2$ fb$^{-1}$) for mSUGRA beyond the current existing bounds from
LEP2 experiments. More recently, the authors of
Ref.~\cite{barger} found that some reach might be recovered in the three
lepton channel if one can use leptons with $p_T$ as small as 5~GeV. 
Potentially worrisome physics backgrounds
from heavy flavor production (which are very effectively reduced with harder
lepton cuts), as well as instrumental
backgrounds from lepton mis-identification
are thought to be under control \cite{kamon}. Nevertheless it is then
natural to ask if the CERN LHC reach for mSUGRA at large $\tan\beta$ is
also diminished, and if in fact SUSY could hide from LHC searches if the
parameter $\tan\beta$ happens to be large.

We have several goals for this paper.
\begin{enumerate}
\item We wish to establish the range of parameter space of the mSUGRA model
that can be probed by CERN LHC experiments. In particular, is the reach of 
the LHC sufficient to discover or rule our mSUGRA at large $\tan\beta$, or 
could mSUGRA effectively hide from SUSY searches? This issue is addressed 
in Sec.~2.
\item If a SUSY signal can be established at large $\tan\beta$, is it
still possible for LHC experiments to make precision measurements of
(differences of) sparticle masses and model parameters? In Sec.~3, we perform a
case study for the parameter space point ($m_0,m_{1/2},A_0,\tan\beta
)= (200\ {\rm GeV},200\ {\rm GeV},0,45$), where $\tw_1$ and $\tz_2$
decay almost exclusively to $\tau$ leptons, to answer this
question. For this case, we find that an edge is reconstructable in
the $\tau^+\tau^-$ invariant mass distribution which gives information
on $m_{\tz_2}-m_{\tz_1}$. Combining the $\tau$'s with jets to form an
invariant mass can also give an estimate of the mass of any squarks produced
in SUSY events.
\item In Sec.~4, we examine the extent to which the $\tau$ lepton
multiplicity exceeds the electron multiplicity in SUSY events if
$\tan\beta$ is large. We discuss various complications for such a
measurement, and point out that $\tau$ signals could provide a novel
diagnostic for SUSY analysis, and that these signals could provide an
alternative handle on the magnitude of $\tan\beta$.


\end{enumerate}
We end with a summary of our results and some general remarks
in Sec.~5.

\section{The Reach of the CERN LHC for mSUGRA at large $\tan\beta$}

We evaluate the mSUGRA signal using the ISAJET 7.37 event generator
program, which is described in more detail in Ref.~\cite{isajet}. We
use the same toy detector simulation as in Ref.~\cite{baer}.

For $m_{\tg},\ m_{\tq} \alt 1$~TeV, $\tg\tg$, $\tg\tq$ and $\tq\tq$
production are the dominant sources of SUSY events at the LHC. These
production mechanisms, together with $\tg$ and $\tq$ cascade decays,
naturally lead to events with $n$-leptons $+ \ m$-jets $+ \ \eslt$,
where typically $n=0\hbox{ -- }4$ and $m \geq 2$. In our simulation,
we generate all SUSY processes using ISAJET. However, our cuts are
designed to selectively pick out gluino and squark events, whose
characteristics are high transverse momentum jets and large missing
transverse energy. Furthermore, the $p_T$ of the primary jets from
gluinos as well as the $\eslt$ are expected to scale with
$m_{\tg}$. In contrast, the momenta of leptons, produced far down in
the cascade decay chain from chargino and neutralino daughters, will
in general be much softer than the jets and $\eslt$, which can be
produced in the first step of the cascade decay. Thus, following
Refs.~\cite{baer} for the multilepton plus multijet signals for SUSY,
we vary the missing-energy and jet $E_T$ cuts using a parameter
$E_T^c$ but fix the lepton cuts:
\begin{itemize}
\item jet multiplicity, $n_{\rm jet} \geq 2$ (with $E_T({\rm jet}) >
100$~GeV),

\item transverse sphericity $S_T > 0.2$,

\item $E_T(j_1), \  E_T(j_2) \ > \ E_T^c$ and $\eslt > E_T^c$.
\end{itemize}

We classify the events by the multiplicity of {\it isolated} leptons,
and in the case of dilepton events, we also distinguish between the
opposite sign (OS) and same sign (SS) samples as these could have
substantially different origins. For the leptons we require,
\begin{itemize}
\item $p_T(\ell) > 20$~GeV ($\ell=e$ or $\mu$) and $M_T(\ell,\eslt) > 100$~GeV
for the $1\ell$ signal, and
\item $p_T(\ell_1,\ell_2) > 20$~GeV for $n=2,3,\ldots$ lepton signals. 
\end{itemize}

The SM background to the various multilepton plus multijet plus
$\eslt$ signal events was calculated in Refs.~\cite{baer} for the
processes $t\bar{t}$ production, $W$ or $Z$ plus jets production,
$WW$, $ZZ$ and $WZ$ production and QCD jet production (where leptons
can arise from decays of heavy flavors produced directly or via gluon
splitting). We use these numerical results for our background
estimates. For each point in mSUGRA parameter space, we require that,
for 10 fb$^{-1}$ of integrated luminosity, the number of signal events
$S$ exceed $5\sqrt{B}$, where $B$ is the number of background events,
for {\it some} value of the cut parameter $E_T^c$. We
also require $S\ge 0.2 B$, and further, that $S\ge 5$ as the minimum number of
events in 10 fb$^{-1}$.

Our results for the reach of the LHC are presented in Fig.~\ref{FIG1}
in the $m_0\ vs.\ m_{1/2}$ plane, for $A_0=0$, $\mu >0$ and {\it a})
$\tan\beta =2$, {\it b}) $\tan\beta =20$, {\it c}) $\tan\beta =35$ and
{\it d}) $\tan\beta =45$. We take $m_t=170$ GeV. The bricked regions
are excluded by lack of appropriate radiative electroweak symmmetry
breaking, or (for $m_0^2 \ll m^2_{1/2}$) if the lightest neutralino
$\tz_1$ is not the lightest SUSY particle (the LSP). The shaded
region is excluded by experimental searches for supersymmetry, and
mainly corresponds to the LEP2 bound $m_{\tw_1}>85$ GeV, or the SUSY
translation of the LEP2 bound that $m_{H}>88$ GeV for a SM Higgs
boson. Mass contours for a 1000 GeV gluino and 1000 GeV first
generation squark are shown to orient the reader.

The results shown in Fig.~\ref{FIG1}{\it a} for $\tan\beta =2$ are
updated versions of similar results presented in Ref.~\cite{baer}, and
are useful for comparison with the higher $\tan\beta$ cases shown in
frames {\it b}), {\it c}) and {\it d}). The largest reach is generally
obtained in the single isolated lepton plus jets plus $\eslt$ channel
(labelled $1\ell$) or in the jets plus $\eslt$ channel with no isolated
leptons (labelled $\eslt$).  For 10 fb$^{-1}$ of integrated luminosity,
values of $m_{\tg}=2300$ GeV (1600 GeV) can be probed for small (large)
values of $m_0$.  Contours for jets plus $\eslt$ plus two opposite sign
isolated leptons (labelled OS), or two same sign leptons (labelled SS)
or three isolated leptons (labelled $3\ell$) are also shown. Each of
these multilepton channels also gives a significant reach for mSUGRA, so
that for much of the parameter space shown, a SUSY signal ought to be
visible in several different channels.

In the case of reach projections for the Fermilab Tevatron, as
$\tan\beta$ increased, it became more difficult to obtain high $p_T$
isolated leptons since chargino and neutralino branchings to $\tau$'s
and $b$'s increased at the expense of $e$'s and $\mu$'s. Consequently,
as $\tan\beta$ increased, the Fermilab Tevatron reach for mSUGRA
decreased, and in fact for $\tan\beta =45$, there was {\it no reach} for
Tevatron Run 2 (2 fb$^{-1}$) beyond the region already excluded by
LEP2\cite{ltanbtev}. As mentioned, if it is possible to use softer cuts
on the leptons, the situation might be somewhat ameliorated \cite{barger}.

The corresponding situation for the CERN LHC is shown in frames {\it
b}), {\it c}) and {\it d}) of Fig.~\ref{FIG1}. For large $\tan\beta$,
we see first of all that the theoretically excluded region increases
substantially at large $m_0$. This region actually depends somewhat
sensitively on the assumed value of the top mass (and on which higher
order corrections are included in the program used). This excluded
region also increases at low $m_0$ and large $m_{1/2}$ when
$m_{\ttau_1}$ becomes lighter than $m_{\tz_1}$.

Next, as $\tan\beta$ increases from 2 to 20, 35 and 45, we see that
the ultimate reach for mSUGRA only decreases slightly in the $1\ell$
and $\eslt$ channels, and only for low values of $m_0$. 
If $m_0$ is large, for $m_{1/2}$ close to the LHC reach in frame {\it
a}), charginos and neutralinos mainly decay to real $W$, $Z$ and Higgs
bosons, and leptonic signals from these decays, and hence the LHC reach, 
are only weakly dependent on $\tan\beta$. For low values of $m_0$,
however, the decays $\tz_2\to\ttau_1\tau$ and $\tw_1\to
\ttau_1\nu$ (bars are omitted), and possibly, those to other sleptons are
also kinematically accessible. The decays to stau dominate for high
values of $\tan\beta$, while for low $\tan\beta$ and low
$m_0$, $\tz_2\to \tz_1 h$ or $\tell\ell$ and $\tw_1\to W\tz_1$ or
$\tell\nu$ or $\ell\tnu$. It is easier to get hard isolated leptons
from the subsequent $W$ or $\tell$ decays than from a $\ttau_1\to
\tau\tz_1\to\ell\nu\nu\tz_1$ decay, which accounts for the somewhat
higher reach in the $1\ell$ channel at low $\tan\beta$. Similarly, the
reach for mSUGRA in the multilepton channels decreases as $\tan\beta$
increases, but again only for low $m_0$.

Although the LHC reach for mSUGRA is somewhat reduced at large
$\tan\beta$, the contours still lie far beyond parameter space
preference curves due to ``naturalness'' considerations\cite{castano},
which tend to lie below the $m_{\tg},m_{\tq} =1000$ GeV contours. 
It is worth noting that the selection criteria designed to extract the
SUSY signal for the low $\tan\beta$ regime suffice even if $\tan\beta$
is large; {\it i.e.} no new analysis is necessary.
The large reach for mSUGRA at large $\tan\beta$ is due in part to the
large squark and gluino production cross sections, and the fact that
for very large sparticle masses, leptons occuring very far down the
cascade decay chain can still have substantial $p_T$. From these reach
contours, we conclude that it would be difficult for mSUGRA to hide
from detection at the LHC by virtue of having a large value of
$\tan\beta$. This is in sharp contrast to the corresponding situation for
the Fermilab Tevatron $p\bar{p}$ collider \cite{ltanbtev}.

\section{A large $\tan\beta$ mSUGRA model case study}

In Refs.~\cite{snowmass,hinchliffe}, five mSUGRA parameter space
points were adopted for detailed case studies. It was found that in
fact precision measurements of (differences of) SUSY particle masses
and model parameters could in many instances be made at LHC
experiments. We briefly summarize the results of
Refs.~\cite{snowmass,hinchliffe} as follows.
\begin{itemize}
\item A global variable
\begin{equation}
M_{eff}=p_T(j1)+p_T(j2)+p_T(j3)+p_T(j4)+\eslt
\end{equation}
(scalar sum) for events with $\ge 4\hbox{-jets}+\eslt$ was defined. 
Distributions in $M_{eff}$ were shown to be dominated at low values by SM 
backgrounds, but were dominated by the mSUGRA signal at high values of
$M_{eff}$. 
The peak in the $M_{eff}$ distribution scaled with $min(m_{\tg},m_{\tq})$
and provided a good first estimate of the strongly interacting SUSY particle
masses involved in the signal events.

\item Gluino and squark cascade decay events involving $\tz_2$ were
found to be very useful for reconstructing the cascade decay chain whenever
the branching fraction for the decay $\tz_2\to h\tz_1$ or the decay
$\tz_2\to \ell\bar{\ell}\tz_1$ 
is substantial. In the case that $\tz_2\to h\tz_1$ followed
by $h\to b\bar{b}$, the $b\bar{b}$ mass could be reconstructed to yield
$m_h$; then by combining with other hard jets present in the events,
mass estimates could be made of other SUSY particles occurring earlier
in the cascade decay sequence.  In the $\tz_2\to \ell\bar{\ell}\tz_1$
case, the endpoint of the $m(\ell\bar{\ell})$ distribution 
leads to a precise determination of 
$m_{\tz_2}-m_{\tz_1}$ (or if the decay is mediated by an on-shell
slepton, yields information about $m_{\tell}$); 
again, by combining dilepton masses with various jets, other
sparticle mass estimates could be obtained. These reconstructed decay
chains thus yield information of several sparticle masses which can then
be used to constrain the underlying model parameters.

\item Finally, a global fit of event characteristics and/or mass measurements
to various SUSY model parameters could be made. An overconstrained fit 
allowed rejection of many possible SUSY models, while honing in on 
possible choices for underlying model parameters.
\end{itemize}
The mSUGRA parameter space choices made in Ref.~\cite{snowmass,hinchliffe}
were necessarily restricted to values of $\tan\beta \le 10$, since ISAJET
was at the time only valid for that parameter space regime.

We expect that the $M_{eff}$ distribution will continue to yield a
measure of the SUSY mass scale regardless of the magnitude of
$\tan\beta$.  However, any sparticle mass reconstruction strategies
involving $\tz_2$ decays are in need of re-analysis since at large
$\tan\beta$, $\tz_2\to \tau\bar{\tau}\tz_1$ can be the dominant decay
mode of $\tz_2$. For this reason, we select an additional case study
point, ``LHC point 6'', with mSUGRA parameter values
$(m_0,\ m_{1/2},\ A_0,\ \tan\beta ,\ sign(\mu ))= (200,\ 200,\ 0,\ 45,\
-1)$, as suggested in \cite{smtheory}, where mass parameters are in
units of GeV.  In this case, $m_{\tg}=540$ GeV, $m_{\tq}=498-517$ GeV,
$m_{\tb_1}=390$ GeV, $m_{\tz_2}\simeq m_{\tw_1}=152$ GeV, $m_{\tz_1}=81$
GeV, $m_{\ttau_1}=131$ GeV and $m_{\tell_R}=219$ GeV. In this case,
$\tz_2\to\ttau_1\tau$ at 99.8\% and $\tw_1\to \ttau_1\nu_\tau$ at
99.6\%. In addition, $\tg\to\tb_1 b$ occurs at 55\%. The SUSY signal
events in this case are expected to be rich in $b$-jets and
$\tau$-leptons.

A signal sample of 500 k events (corresponding to an integrated
luminosity of about 5~fb$^{-1}$) was generated, along with 250 k event
background samples each for $W+$jets, $Z+$jets, $t\bar{t}$ 
production as in Ref.~\cite{hinchliffe}. QCD backgrounds are expected to
be small after basic selection cuts discussed below. Hadronic $\tau$'s were
found using generator information rather than selecting narrow jets.  In
addition, hadronic $\tau$'s were required to have visible $p_T>20$ GeV
and $|\eta |<2.5$. We used the CDF rate \footnote{The CDF analysis also
involves other cuts that we have not applied to the data.} for jets to
fake $\tau$'s, namely, 0.5\% at $p_T(jet)=20$ GeV and 0.1\% for
$p_T(jet)>50$ GeV, with a linear interpolation in
between\cite{cdftau}. The following standard cuts were made:
\begin{itemize}
\item at least four jets were required (using the ISAJET GETJET routine)
with $p_T(j_1)>100$ GeV and $p_T(j_{2,3,4})>50$ GeV (leptons and taus
are counted as jets),
\item $\eslt >100$ GeV,
\item transverse sphericity $S_T>0.2$,
\item $M_{eff}>500$ GeV.
\end{itemize}
Efficiencies of 60\% for $b$-tagging and 90\% for lepton and
(hadronically decaying) $\tau$
identification were assumed. (The $\tau$ efficiency is too optimistic, but
studying it requires more than a toy detector simulation.) With these
cuts, the event sample was already dominated by signal so that errors
in the mis-identification rate of $b$'s and $\tau$'s are not expected
to be a problem.

The distribution in $M_{eff}$ is shown in Fig.~\ref{MEFF}. The signal
is shown by the solid histogram, while background is shown by the
dashed histogram. The signal easily dominates background as expected
for large values of $M_{eff} \geq 650$ GeV. At low $\tan\beta$ the
ratio of $M_{eff}$ where signal just exceeds background to $M_{SUSY}$ is
noted\cite{hinchliffe} to be $\sim 1.5-1.6$, and provides an estimate
of $M_{SUSY}$. In this case, the ratio is $\sim 1.3$, which is due
mainly to the larger mass splittings of squarks ($m_{\tb_1},\
m_{\tst_1} \ll m_{\tilde u_{L,R}}$) at large $\tan\beta$.

For events with at least two hadronic taus, we plot
in Fig.~\ref{FIG2} the visible $\tau\tau$ mass for the two highest $p_T$
tau leptons.
The distribution shown exhibits an edge near the endpoint for
$\tz_2\to\tau\bar{\tau}\tz_1$, but the signal to SUSY
background ratio is poor
since a large fraction of the mass is lost to neutrinos; {\it i.e.}
a substantial number of SUSY events with the ``real mass'' beyond the end
point appear below the end point because of the mass carried off in neutrinos. 
The rate is very much 
larger than the SM background shown as the hatched histogram, 
so observing a signal in this 
distribution would be trivial. The end point of the neutralino decay
appears as a kink in the $m_{\tau\tau}$ distribution.
The visible $\tau\tau$ mass is calculated using 
generator information and so does not include the effects of
calorimeter and/or tracker resolution.

A more faithful $\tau\tau$ mass distribution can be obtained by
selecting multiparticle hadronic $\tau$ decays with a visible mass close
to the $\tau$ mass to reduce the mass carried off by neutrinos. The
$\tau^+\tau^-$ mass distribution for 3-prong $\tau$ decays from the
signal and SM background is shown in Fig.~\ref{FIG3} for events with
exactly two opposite sign hadronic $\tau$'s and no additional isolated
leptons.  The requirement of exactly two $\tau$'s was imposed to remove
combinatorial background. There is very little real SM $\tau$ background
after cuts, but there still is a substantial contamination from other
SUSY sources. Since most of the SUSY events contain at least one
gluino, which is a Majorana fermion and has equal branching ratios to
$\tau^+$ and $\tau^-$, the background from two independent chargino
decays can be estimated from the $\tau^\pm\tau^\pm$ distribution, which
is also shown as the dashed histogram in Fig.~\ref{FIG3}.

The subtracted distribution, shown in Fig.~\ref{FIG4}, has a clear low mass
enhancement with an endpoint near but slightly below the limit for
two body decays neglecting the $\tau$ mass:
\begin{equation}
M_{max}=m_{\tz_2}\sqrt{1-\frac{m^2_{\ttau_1}}{m^2_{\tz_2}}}
\sqrt{1-\frac{m^2_{\tz_1}}{m^2_{\ttau_1}}} = 60.6\ {\rm GeV}.
\end{equation}
The contribution to this distribution of events with exactly one $\tz_2$
and no $\tw_1$ is shown as the dashed curve in the figure and clearly
accounts for most of the low mass enhancement. There is also a
contribution from events containing at least one $\tz_3$ or $\tz_4$, the
dash-dot curve in Fig.~\ref{FIG4}, which accounts for the excess of
$\tau^+\tau^-$ pairs at higher mass. Many channels contribute to these
decays; the branching ratio for $\tz_4\to\ttau_1\tau$ is only 8.8\%, so
the distribution appears to end before the kinematic limit,
\begin{equation}
M_{max}=m_{\tz_4}\sqrt{1-\frac{m^2_{\ttau_1}}{m^2_{\tz_4}}}
\sqrt{1-\frac{m^2_{\tz_1}}{m^2_{\ttau_1}}} =216.0\ {\rm GeV}.
\end{equation}
Thus, from these distributions, it should be possible to extract information
not only on $m_{\tz_2}-m_{\tz_1}$, but perhaps also information on heavier 
neutralino masses as well, although this will probably be very
difficult. 

We should point out that while focussing on taus with three
charged prong decays indeed gives us a truer di-tau mass distribution, 
it also leads to a reduction in event rate by more than an order of
magnitude. We have not attempted to examine
whether the end point is better determined from this ``truer''
distribution or from the kink in the distribution in Fig.~\ref{FIG2}
which includes an order of
magnitude larger data sample. 
Our study here should be regarded as a first look at the sort of
measurements that might be possible when charginos and neutralinos
dominantly decay to taus.

Most of the $\tz_2$'s originate from squark cascade decays in this case study.
It is then interesting to see if an estimate can be made of the squark 
masses as well as the neutralino masses. In Fig.~\ref{FIG5}, we have
required events with exactly two $\tau$'s, each decaying into three
prongs, and then have constructed the invariant mass of the $\tau$-pair with 
each of the two fastest jets; finally, we plot the minimum of these two
masses. Since the hardest jets typically come from 
$\tq\to q\tz_i$ decay, this distribution should be approximately bounded
by the squark mass. This is verified in Fig.~\ref{FIG5}, where the bulk of
the $m_{\tau\tau j}$ distribution is in fact bounded by 
$m_{\tq}\simeq 500$ GeV.
 
A similar calculation can be performed using only identified $b$-jets, to try
to extract the $\tb_1$ mass from $\tb_1\to b\tz_2$ decays. In Fig.~\ref{FIG6},
we plot the distribution for $M_{\tau\tau b}$ using again the smaller of the 
two mass combinations. The bulk of the distribution is bounded by
$m_{\tb_1}=390$~GeV, although a tail extends to higher mass values. 
This is due in part to contributions from $\tb_2$ decays, where 
$m_{\tb_2}=480$~GeV. It might also be interesting to see whether it is
similarly possible to isolate the decay chain $\tg \to b\tb_1 \to bb\tz_2 \to
bb\tau\tau\tz_1$ by looking at the $M_{bb\tau\tau}$ which should be
bounded by $m_{\tg}$.

\section{Lepton non-universality at large $\tan\beta$}

Over a significant portion of mSUGRA parameter space it is expected
that the multiplicity of $\tau$ leptons should be enhanced relative to
$e$s or $\mu$s at large $\tan\beta$.  This suggests that if it is
possible to conclusively establish tau
lepton non-universality in SUSY events, we may be able to interprete it
as an indicator of a sizeable tau Yukawa interaction, at least within
the mSUGRA framework. It should, of course, be kept in mind
that within the more general MSSM framework, SUSY events may exhibit
lepton non-universality even if Yukawa couplings are negligible, as long as
$m_{\ttau_1}$ differs from selectron and smuon masses. The
observation of universality between $e$ and $\mu$ in a SUSY event sample would
strongly tempt us to suggest\cite{ftnote} that an observed tau 
non-universality indeed originates
in a sizeable tau Yukawa coupling.

While the principle is simple, its implementation poses a challenge.
Even in a sample of purely SM events, there should be a superficial
non-universality of $e:\mu :\tau$ simply due to the different acceptance
cuts and efficiencies for identifying each species of lepton. It should
be possible to determine these directly from a data sample rich in SM
events. However, these efficiencies will also change somewhat with the
cuts used to select out SUSY events, but this can presumably be taken
into account, again using the data; {\it e.g.} by using selection cuts
that smoothly interpolate between SM and signal samples.  
Hadronically decaying taus pose yet
another challenge since the visible energy spectrum in their decays, and
hence, the tau detection efficency, is sensitive to their
polarization~\cite{hag}. 
Moreover, for a data sample enriched in New Physics (in our case SUSY)
events, the tau polarization is not known {\it a priori}, but may be
possible to determine from the data.

A complete quantitative analysis of lepton non-universality is beyond
the scope of this study. For one, it would entail a correct simulation of tau
decays (including effects of tau polarization) both for SM backgrounds and the
SUSY sample. These effects are not yet completely included in
ISAJET. Our study should, therefore, be regarded as exploratory, and
simply indicative of the magnitude of asymmetries that we will have at
our disposal for future studies.

To illustrate this with a direct computation, we examine three 
different mSUGRA parameter space points with $m_0=225$ GeV, $m_{1/2}=250$ GeV,
$A_0 =0$, $\mu >0$ and {\it A}) $\tan\beta =2$, {\it B}) $\tan\beta =35$
and {\it C}) $\tan\beta =45$, along with the SM background.
We impose a simple set of cuts:
\begin{itemize}
\item we require 2 jets each with $E_T>E_T^c$ and $\eslt >E_T^c$,
\item we require transverse sphericity $S_T>0.2$,
\item for events with a single isolated lepton, we 
require $M_T(\ell ,\eslt )>100$ GeV ($\ell= e,\ \mu$ or $\tau$), where in the
case of $\tau$ we use the visible $\tau$ energy to construct $M_T$.
\end{itemize}
We show the total signal and background as a function of $E_T^c$ in
Fig.~\ref{NONU}{\it a}. For low values of $E_T^c$, the resulting event
sample is background dominated, while for high $E_T^c$, the sample is
signal dominated. In Fig.~\ref{NONU}{\it b}, we plot the average
lepton multiplicity in the surviving signal plus background event sample of
$e$'s and $\tau$'s, denoted by $ \langle n_e \rangle $ (dashed) and $
\langle n_{\tau} \rangle$ (solid). For case A, with low $\tan\beta =2$,
the $\tau$ multiplicity is roughly constant versus $E_T^c$, and would
correspond to a measured lepton universality after accounting for
acceptances and efficiencies. For this case, the quantity $\langle n_e
\rangle $ decreases somewhat with $E_T^c$ since jetty gluino and squark
cascade decay events are more likely to pass our simple cuts listed
above. For case B with $\tan\beta =35$, $\tw_1\to W\tz_1$ with a
branching fraction of 98\% while
$\tz_2\to \tz_1\tau^+\tau^-$ 27\% of the time, so that some violation of
universality is expected. This is seen in Fig.~\ref{NONU}{\it b} where
the $e$ and $\tau$ multiplicity is nearly that of case A for low
$E_T^c$, while for high $E_T^c$, which is signal dominated, there is a
distinct increase in $\tau$ multiplicity compared to $e$
multiplicity. This can be seen more easily in Fig.~\ref{NONU}{\it c}
where we plot the ratio $\langle n_{\tau} \rangle / \langle n_e \rangle$
versus $E_T^c$. For case C, we have $B(\tw_1\to\ttau_1\nu_{\tau})$ = 93\%
and $B(\tz_2\to \ttau\tau)$ = 99\%, so large deviations from universality
should be expected at high $E_T^c$. For this case, we see in
Fig.~\ref{NONU}{\it b} that in fact $\langle n_{\tau} \rangle $
surpasses $\langle n_e \rangle$ for all $E_T^c > 50$ GeV, and
Fig.~\ref{NONU}{\it c} shows the huge deviations from universality that
would be expected for very large $\tan\beta$ and small $m_0$.

While we recognize that our results should be regarded as
qualitative, we are encouraged to see that the magnitudes
of asymmetries in the three cases are quite different. As more complete
simulations become available, it would be instructive to study the
extent to which $\tau$ leptons may serve as a diagnostic of any new
physics that might be discovered. Such analyses will have to be
interpreted with care since, as we have said, the tau detection
efficiency, and hence the expectation for tau multiplicity, depends on
the unknown polarization (which may be possible to measure) in the new
physics sample. Well-defined frameworks such as mSUGRA would, however,
make unambiguous predictions for $\langle n_{\tau} \rangle $, so that
this measurement could serve as an independent test, and possibly even
provide a measure of $\tan\beta$ (especially if it happens to be
large). Indeed in the future, it may prove worthwhile to examine the
multiplicity of taus separately in various event toplogy samples
($1\ell, \ell^+\ell^-, \ell^{\pm}\ell^{\pm}, 3\ell$ {\it etc.}) since
these generally have different SUSY origins.

\section{Summary and Conclusions}

In this paper we have studied SUSY signals at the CERN LHC collider
as predicted by the minimal supergravity model for large values of the
parameter $\tan\beta$. We found that increasing this parameter to
values near its upper bound has little impact on the SUSY discovery
reach of the LHC, in stark contrast to the Tevatron, whose reach is
greatly diminished in this region of parameter space. The main reason
for this difference is that increasing $\tan\beta$ can change the
qualitative pattern of neutralino and chargino decays only for
relatively small sparticle masses, where their decays into real $W$ and $Z$
bosons are kinematically disallowed. Even though in this region of
parameter space the efficiency for detecting SUSY through events
containing hard leptons is low also at the LHC, the huge event rate
guarantees that SUSY will still be seen in several different channels.
Once decays into real gauge or Higgs bosons become possible, decay
patterns become less sensitive to $\tan\beta$; in particular, except
when $m_0$ is very small, many
hard leptons now come from the decay of on--shell $W$ and $Z$ bosons,
independent of the value of $\tan\beta$.

Apart from discovering SUSY, one would also like to determine its
parameters, so as to eventually pin down the supersymmetric model that
describes nature at a more fundamental level. Most previous studies that
attempted to reconstruct some (differences of) SUSY masses at the LHC
used events with hard isolated leptons. Unfortunately, for large values
of $\tan\beta$ and not too heavy sparticles SUSY events are expected to
contain $\tau$ leptons rather than electrons or muons. While
leptonically decaying $\tau$'s still produce sufficiently many hard
electrons and muons to ensure that SUSY will be discovered, the presence
of many additional neutrinos would make mass reconstruction using
electrons and muons all but impossible in this part of parameter
space. In Sec.~3 we instead used hadronic 3--prong decays of $\tau$'s to
determine the difference between the masses of the lightest and
next--to--lightest neutralino; we focussed on this decay mode since here
the simultaneously produced $\nu_\tau$s (which smear out the end point)
are forced to be relatively soft, which minimizes their impact on
kinematic event reconstruction. While the precision of this measurement
will be worse than that of the analogous measurement based on $e^+ e^-$
pairs at smaller $\tan\beta$, it should still be sufficient to greatly
constrain the SUSY model. We also showed how combining $\tau$ pairs with
a hard jet might yield information about the overall scale of the squark
masses or, if this jet contains a tagged $b$, the mass of bottom
squarks.

Finally, the presence of many $\tau$ leptons also offers new 
opportunities to glean information about the underlying SUSY parameters.
As a first attempt in that direction we studied in Sec.~4 the
violation of lepton universality that can be expected in SUSY events
if $\tan\beta$ is large. In order to make this fully quantitative,
a careful analysis of the different detection efficiencies for the
three flavors of leptons is mandatory, which requires detailed
understanding about the performance of LHC detectors as well as detailed
simulation of $\tau$ decays. Here we instead
simply showed corresponding results for event samples dominated by
Standard Model contributions, where lepton universality is known to
hold to very good approximation; this serves as a normalization for
other event samples dominated by SUSY contributions. Some caution is
advised when interpreting these results. For example, the average
$\tau$ polarization is expected to change when going from the SM--dominated
sample to the SUSY--dominated sample; this will change the $E_T$
spectrum of the visible $\tau$ decay products, and hence their
detection efficiency. However, our results clearly show that at least
for the extreme case where $\tw_1$ and $\tz_2$ almost exclusively decay
into real $\tilde{\tau}_1$ sleptons, a gross violation of lepton
universality should be expected.

One can envision other, more ambitious studies of SUSY events
containing hadronically decaying $\tau$ leptons. For example, a
comparison of the visible spectra of events with 1--prong and 3--prong
$\tau$ decays should allow to determine the $\tau$ polarization, which
in turn would yield information on the $\tilde{\tau}_L -
\tilde{\tau}_R$ mixing angle, as well as the decomposition of the
$\tw_1$ and/or $\tz_2$ mass eigenstates in terms of gaugino and higgsino
current states. Copious production of $\tau$ leptons, a bane for SUSY
searches at the Tevatron, could therefore well turn out to be a boon
for the LHC.

%
\acknowledgments
This research was supported in part by the U.~S. Department of Energy
under contract numbers DE-FG02-97ER41022, DE-AC02-98CH10886, and
DE-FG-03-94ER40833. 
%
%

\newpage
%
%

\iftightenlines\else\newpage\fi
\iftightenlines\global\firstfigfalse\fi
\def\dofig#1#2{\epsfxsize=#1\centerline{\epsfbox{#2}}}



\begin{figure}
\dofig{6in}{fig1.ai}
\caption[]{
A plot of the reach of the CERN LHC for various $n$-lepton plus multijet
plus missing $E_T$ events from mSUGRA in the $m_0\ vs.\ m_{1/2}$ plane
for $A_0=0$, $\mu >0$ and {\it a}) $\tan\beta =2$, {\it b}) $\tan\beta =20$, 
{\it c}) $\tan\beta =35$ and {\it d}) $\tan\beta =45$.
We take $m_t=170$ GeV.}
\label{FIG1}
\end{figure}
\begin{figure}
\dofig{5in}{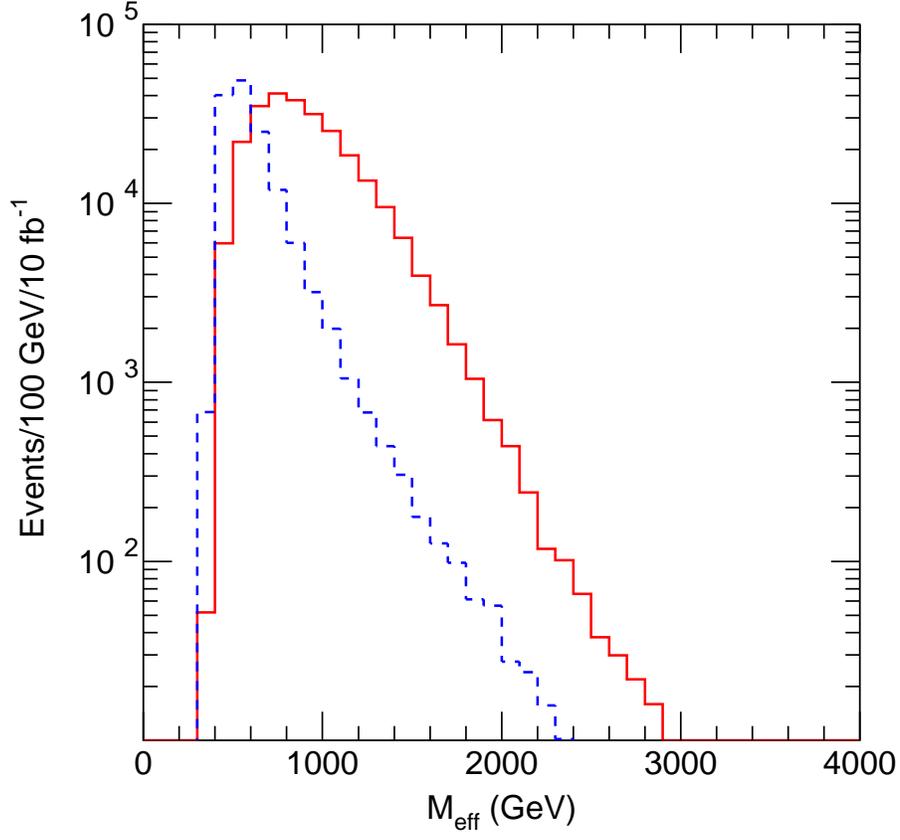}
\caption[]{Distribution in $M_{eff}$ for the case study of Sec.~3. 
The signal is the solid histogram, while the dashed histogram represents
background.}
\label{MEFF}
\end{figure}
\begin{figure}
\dofig{5in}{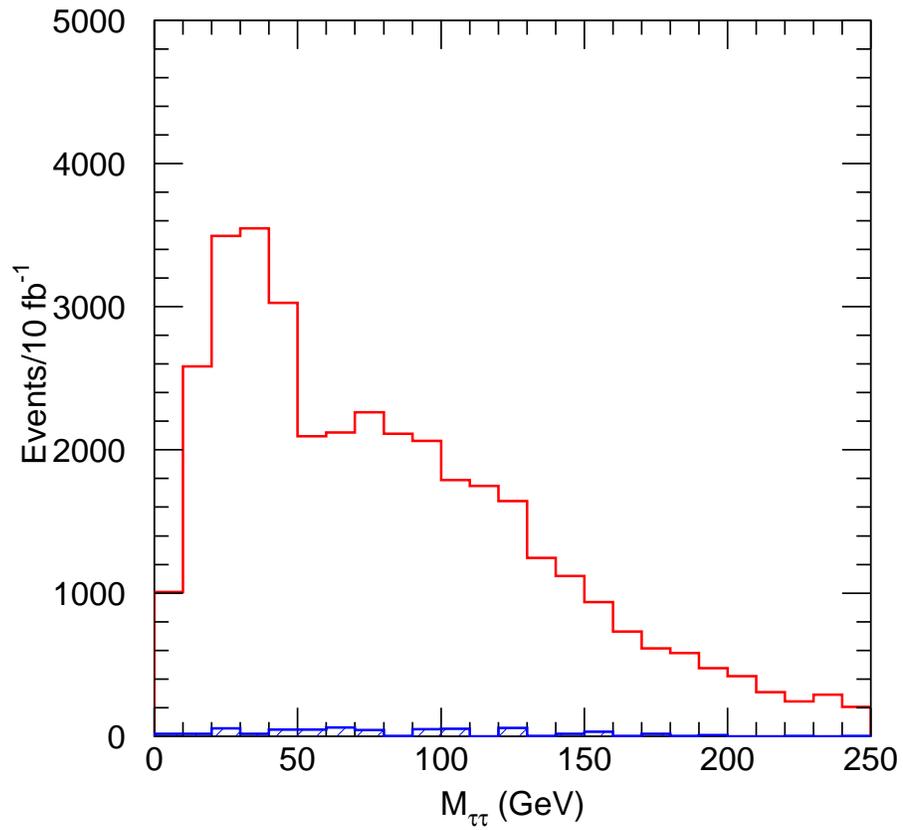}
\caption[]{Visible $\tau\tau$ mass for all hadronic decays.}
\label{FIG2}
\end{figure}
\begin{figure}
\dofig{5in}{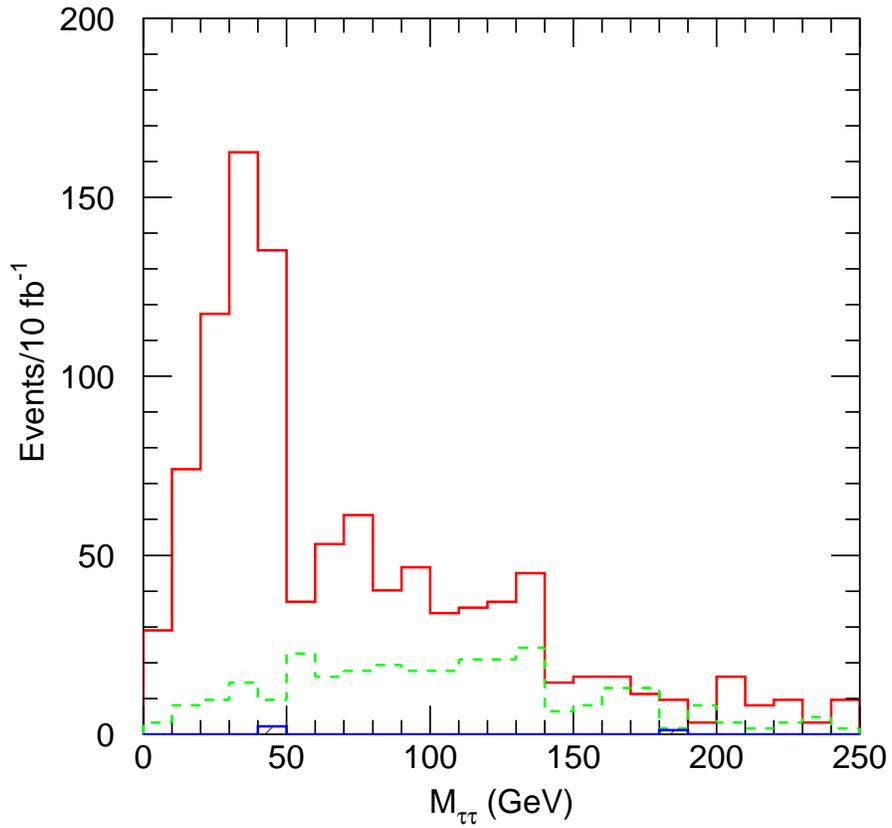}
\caption[]{Visible $\tau\tau$ mass with 3-prong decays (solid), 
SUSY background estimate from $\tau^\pm\tau^\pm$ (dashed), 
and SM background (shaded).}
\label{FIG3}
\end{figure}
\begin{figure}
\dofig{5in}{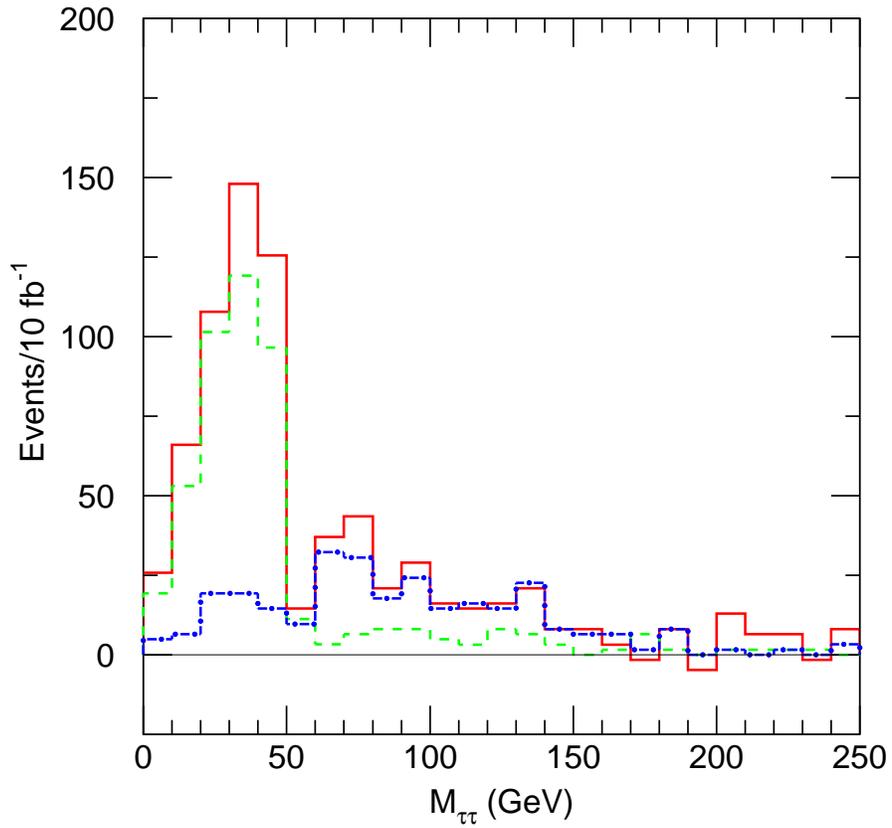}
\caption[]{Visible mass distribution for the difference of
$\tau^+\tau^-$ and $\tau^\pm\tau^\pm$ (solid). The contribution
of $\tz_2$ events (dashed), and $\tz_3 +\tz_4$ events (dashed-dotted)
are also shown.}
\label{FIG4}
\end{figure}
\begin{figure}
\dofig{5in}{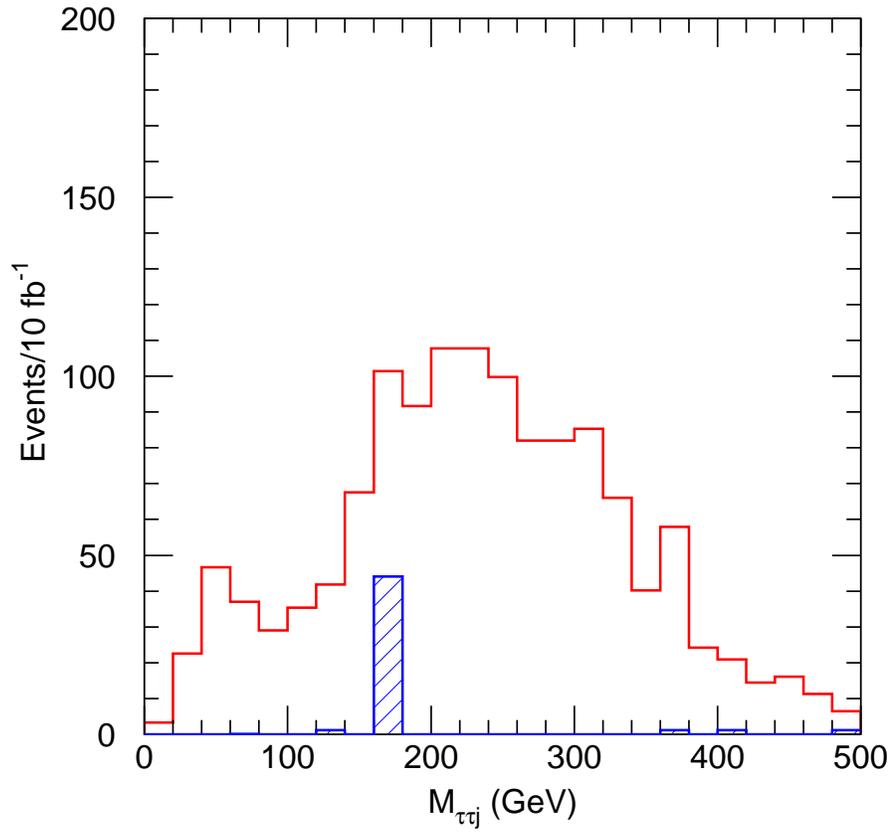}
\caption[]{Visible $\tau\tau -jet$ mass distribution for the smaller of
two combinations. The shaded histogram is the SM background.}
\label{FIG5}
\end{figure}
\begin{figure}
\dofig{5in}{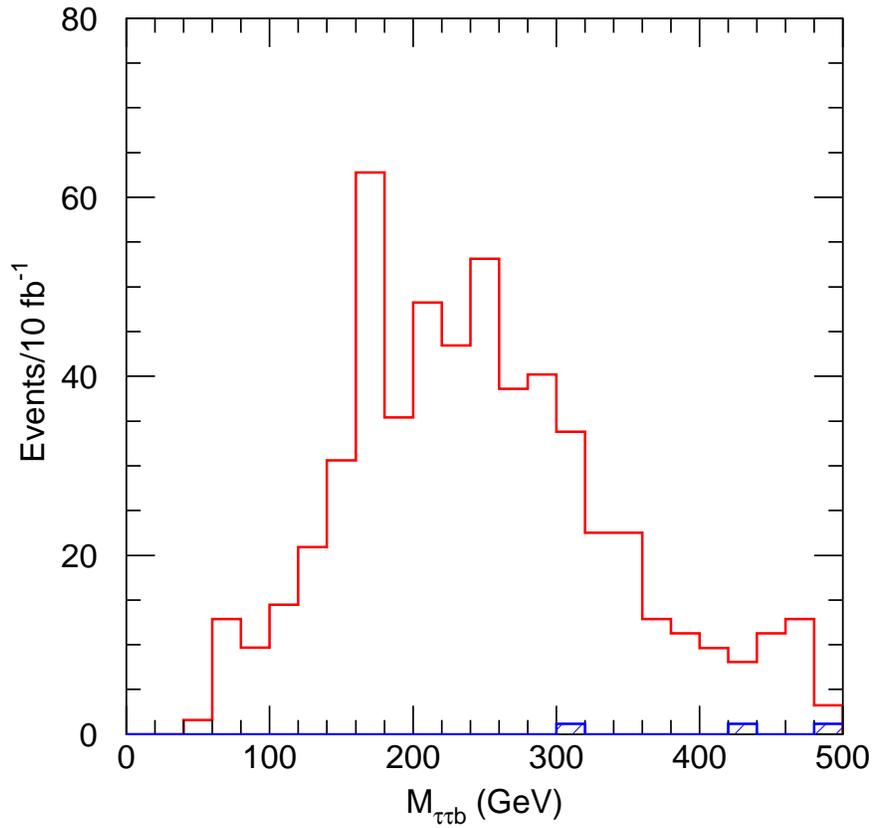}
\caption[]{Visible $\tau\tau b$ mass distribution for the smaller
of two possible combinations. The shaded histogram is the SM background.}
\label{FIG6}
\end{figure}
\begin{figure}
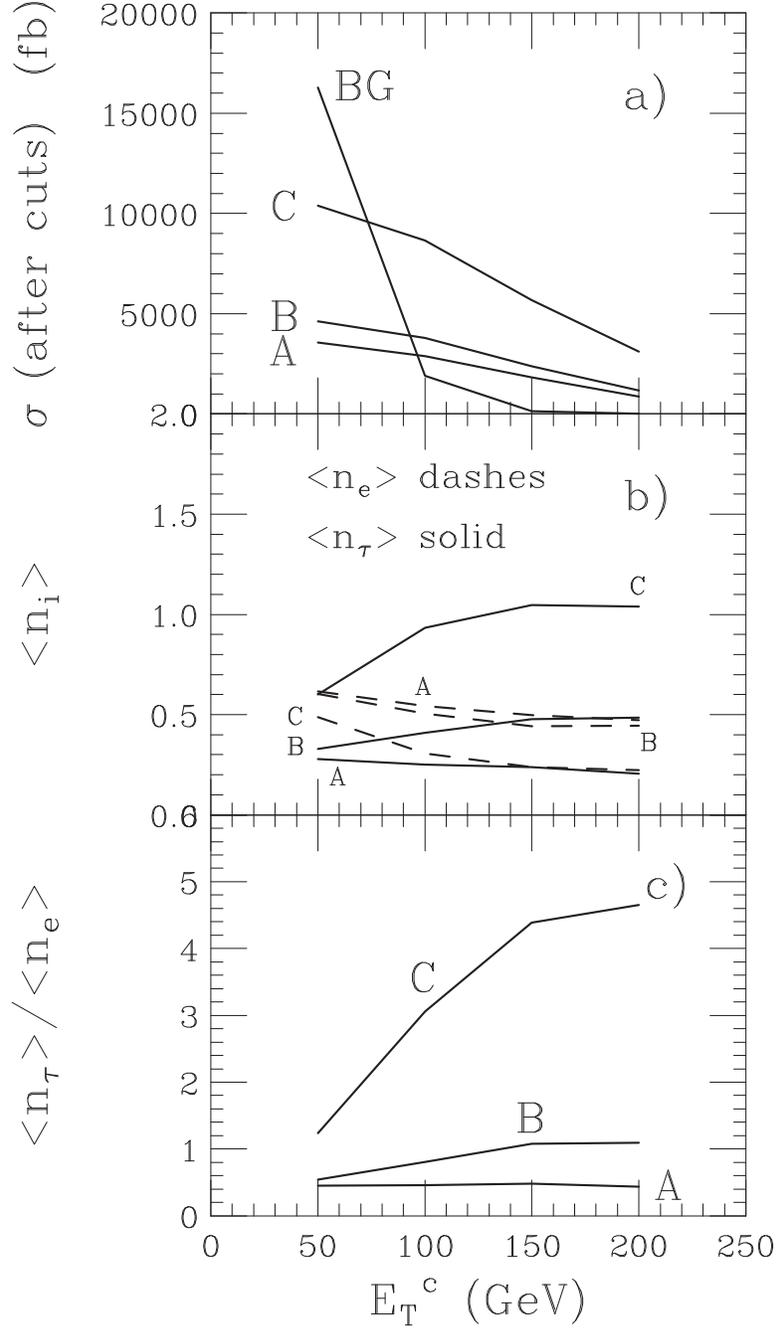

\dofig{4in}{nonu.ai}
\medskip
\caption[]{A plot of {\it a}) background (BG) and 3 signal cases A, B and C
given in the text after modest cuts. In {\it b}), we plot the average
$e$ and $\tau$ multiplicities in the signal plus background sample of events 
that survive cuts. In {\it c}), the ratio of $\tau$ to $e$
multiplicities is plotted versus $E_T^c$ for the three cases of
signal plus background.}
\label{NONU}
\end{figure}


\begin{references}
\bibitem{ftnote} Even this does not rigorously follow in a 
SUSY GUT \cite{bh} because
$\ttau_R$ resides in the same GUT multiplet as the $t$-squarks. Its mass,
therefore, undergoes substantial evolution due to the large top Yukawa
coupling if the scalar mass unification scale $M_X$ is significantly
larger than $M_{GUT}$. Thus $m_{\ttau_R}$ may differ from the selectron
mass, even when tau Yukawa coupling effects are small.
\end{references}
\end{document}